\documentclass{article}
\usepackage[T1]{fontenc} 
\usepackage[utf8]{inputenc} 
\usepackage{ismir,amsmath,amssymb,cite,url}
\usepackage{multicol,multirow}
\usepackage{graphicx}

\usepackage{color}

\usepackage[firstpage]{draftwatermark}
\definecolor{lightgray}{rgb}{0.9,0.9,0.9}
\definecolor{darkgray}{rgb}{0.4,0.4,0.4}
\SetWatermarkFontSize{12pt}
\SetWatermarkScale{1.1}
\SetWatermarkAngle{90}
\SetWatermarkHorCenter{202mm}
\SetWatermarkVerCenter{170mm}
\SetWatermarkColor{darkgray}
\SetWatermarkText{Late-Breaking / Demo Session Extended Abstract, ISMIR 2024 Conference}

\def\lbd{}	

\title{Boundary Regression for Leitmotif Detection in Music Audio}

\twoauthors
 {Sihun Lee} {Sogang University \\ \texttt{sihunlee@sogang.ac.kr}}
 {Dasaem Jeong} {Sogang University \\ \texttt{dasaemj@sogang.ac.kr}}

\def\authorname{S. Lee and D. Jeong}

\begin{document}

\maketitle
\begin{abstract}
Leitmotifs are musical phrases that are reprised in various forms throughout a piece. Due to diverse variations and instrumentation, detecting the occurrence of leitmotifs from audio recordings is a highly challenging task. Leitmotif detection may be handled as a subcategory of audio event detection, where leitmotif activity is predicted at the frame level. However, as leitmotifs embody distinct, coherent musical structures, a more holistic approach akin to bounding box regression in visual object detection can be helpful. This method captures the entirety of a motif rather than fragmenting it into individual frames, thereby preserving its musical integrity and producing more useful predictions. We present our experimental results on tackling leitmotif detection as a boundary regression task.

\end{abstract}
\section{Introduction}\label{sec:introduction}
In the \textit{Musikdrama} of the German composer Richard Wagner, leitmotifs are important musical devices that form compositional structure and convey drama. As a short musical phrase, each leitmotif may represent a character, an action, or an idea, and is reprised throughout the multiple hours ---16 in the case of the \textit{Ring} cycle--- of drama, in diverse variations and instrumentation. Recognizing instances of leitmotifs is regarded as an important aspect of the listening experience, and therefore many guides to Wagner's work specify the sections in which leitmotifs appear~\cite{dreyfus2021lohengrin, olson1998interactive}.

Automatic detection of leitmotifs from audio recordings has only been tackled in a limited number of works, notably by Krause et al.~\cite{krause2020classifying, krause2021tismir}. They approached leitmotif activity detection as a subcategory of sound event detection, where a model has to predict the presence of all leitmotif classes for each time frame of audio representation.

Frame-wise prediction can be suitable for the detection of environmental noise(e.g. vehicle traffic, vacuum cleaner) or sound events(e.g. gunshots, dog barking). These types of sounds can be recognized even with very short segments; the noise of a vacuum cleaner, for example, can be detected from a few frames of audio spectrogram. Leitmotifs, on the other hand, often require more context. 
To form a recognizable leitmotif, several consecutive melodic intervals or chords, who may each appear fairly commonly in other contexts, need to be grouped in a specific rhythm. In this regard, it may be more appropriate to represent a leitmotif instance as a single grouped event, rather than as a stream of frame activations.

Moreover, for theoretical and practical use, a precise frame-wise prediction of leitmotif activity may be redundant; predicting the \textit{general time boundary} of an occurrence would be more than sufficient, and could better reflect the malleable nature of leitmotifs. Thus, we suggest that it may be beneficial to approach this task as one similar to image bounding box regression.

Bounding box regression for sound event detection has also shown promising results in recent DCASE challenges~\cite{ebbers2024sound}. In this paper, we present our results on adapting a such an approach for the challenging task of leitmotif activity detection.

\section{Leitmotif Boundary Regression}

We take inspiration from the \textit{YOLO} family of models\cite{redmon2016you, redmon2017yolo9000} in computer vision and build a CNN-based network that outputs a set of boundary predictions for a given audio clip. We apply constant-Q transform (CQT) to incoming 15-second clips with 12 bins per octave and a hop length of 512 samples. The resulting CQT of 646 frames is then fed to a stack of convolution layers, which reduces the frequency dimension to 1 and the time dimension to 11, effectively dividing the audio into 11 grids that each output separate predictions.

Each prediction consists of three values \((p, x, w)\) and a probability distribution for all leitmotif classes, where \(p\in[0, 1]\) is the confidence score, \(x\in[0, 1]\) is the center position of the boundary relative to the grid, and \(w\in\mathbb{R}\) is the width scale of the boundary. Instead of directly predicting the width of boundaries, we utilize anchors \(A = \{ a_1, \dots, a_n \}\), which are predetermined boundary lengths, and take \(a_{n}\times \exp(w)\) to get the final width prediction for the \(n\)-th anchor. Thus, for \(C\) leitmotif classes, the output of the model is a tensor of shape \(n \times 11 \times (3 + C)\), where 11 is the number of grids and \(3 + C\) denotes $(p, x, w)$ and the logits for each class. During experiment, we use \(n = 3\) and \(C = 13\), and use K-means clustering on the leitmotif activity boundaries in the dataset to determine the size of the anchors.

\section{Experiments and Results}
\subsection{Dataset}
We use the Wagner Ring leitmotif dataset published by Krause et al.~\cite{krause2021tismir} and acquire the corresponding audio recordings, which consist of 13 aligned performances and 143 files. We separate the dataset in a "version split", where the entirety of a single recording version is included in one of the split groups. 
Among the 13 performances, we use 10 for training, 2 for validation, and 1 for test, and among the 20 annotated leitmotifs, we use the better-represented 15. This results in a total of 28,820 leitmotif instance samples in the training set. We incrementally divide the audio recordings into 15-second clips with a 14-second overlap, and create ground truth tensors for boundary regression by assigning each leitmotif instance to the best anchor and grid based on Intersection over Union (IoU).

\begin{table}[]
    \centering
    \begin{tabular}{lllllll}
        \cline{1-3} \cline{5-7}
        \multicolumn{3}{c}{Proposed Model} &  & \multicolumn{3}{c}{Baseline Model} \\ \cline{1-3} \cline{5-7} 
        \multicolumn{2}{l|}{\# params} & 1.97M &  & \multicolumn{2}{l|}{\# params} & 2.19M \\ \cline{1-3} \cline{5-7} 
        \multicolumn{1}{l|}{\multirow{13}{*}{\begin{tabular}[c]{@{}l@{}}Boundary-\\wise\\ F\textsubscript{1}\end{tabular}}} & \multicolumn{1}{l|}{\texttt{Ni}} & 0.83 &  & \multicolumn{1}{l|}{\multirow{13}{*}{\begin{tabular}[c]{@{}l@{}}Frame-\\wise\\ F\textsubscript{1}\end{tabular}}} & \multicolumn{1}{l|}{\texttt{Ni}} & 0.91 \\
        \multicolumn{1}{l|}{} & \multicolumn{1}{l|}{\texttt{Ri}} & 0.86 &  & \multicolumn{1}{l|}{} & \multicolumn{1}{l|}{\texttt{Ri}} & 0.86 \\
        \multicolumn{1}{l|}{} & \multicolumn{1}{l|}{\texttt{NH}} & 0.88 &  & \multicolumn{1}{l|}{} & \multicolumn{1}{l|}{\texttt{NH}} & 0.93 \\
        \multicolumn{1}{l|}{} & \multicolumn{1}{l|}{\texttt{RT}} & 0.78 &  & \multicolumn{1}{l|}{} & \multicolumn{1}{l|}{\texttt{RT}} & 0.89 \\
        \multicolumn{1}{l|}{} & \multicolumn{1}{l|}{\texttt{Wa}} & 0.90 &  & \multicolumn{1}{l|}{} & \multicolumn{1}{l|}{\texttt{Wa}} & 0.97 \\
        \multicolumn{1}{l|}{} & \multicolumn{1}{l|}{\texttt{WL}} & 0.87 &  & \multicolumn{1}{l|}{} & \multicolumn{1}{l|}{\texttt{WL}} & 0.90 \\
        \multicolumn{1}{l|}{} & \multicolumn{1}{l|}{\texttt{Ho}} & 0.84 &  & \multicolumn{1}{l|}{} & \multicolumn{1}{l|}{\texttt{Ho}} & 0.90 \\
        \multicolumn{1}{l|}{} & \multicolumn{1}{l|}{\texttt{Sc}} & 0.88 &  & \multicolumn{1}{l|}{} & \multicolumn{1}{l|}{\texttt{Sc}} & 0.91 \\
        \multicolumn{1}{l|}{} & \multicolumn{1}{l|}{\texttt{WH}} & 0.87 &  & \multicolumn{1}{l|}{} & \multicolumn{1}{l|}{\texttt{WH}} & 0.88 \\
        \multicolumn{1}{l|}{} & \multicolumn{1}{l|}{\texttt{Fe}} & 0.93 &  & \multicolumn{1}{l|}{} & \multicolumn{1}{l|}{\texttt{Fe}} & 0.94 \\
        \multicolumn{1}{l|}{} & \multicolumn{1}{l|}{\texttt{Un}} & 0.90 &  & \multicolumn{1}{l|}{} & \multicolumn{1}{l|}{\texttt{Un}} & 0.92 \\
        \multicolumn{1}{l|}{} & \multicolumn{1}{l|}{\texttt{Si}} & 0.89 &  & \multicolumn{1}{l|}{} & \multicolumn{1}{l|}{\texttt{Si}} & 0.89 \\
        \multicolumn{1}{l|}{} & \multicolumn{1}{l|}{\texttt{Ve}} & 0.88 &  & \multicolumn{1}{l|}{} & \multicolumn{1}{l|}{\texttt{Ve}} & 0.92 \\ \cline{1-3} \cline{5-7} 
        \multicolumn{2}{l|}{mAP} & 0.64 &  & \multicolumn{2}{l}{} &  \\ \cline{1-3}
        \multicolumn{2}{l|}{mAP\textsubscript{50}} & 0.93 &  & \multicolumn{2}{l}{} &  \\ \cline{1-3}
        \multicolumn{2}{l|}{mAP\textsubscript{75}} & 0.73 &  & \multicolumn{2}{l}{} &  \\ \cline{1-3}
        \end{tabular}
    \caption{Evaluation results from the proposed and baseline models. Per-class F\textsubscript{1}-scores are calculated boundary-wise for the proposed model, and frame-wise for the baseline model.}
    \label{tab:results}
\end{table}

\begin{figure}
 \centerline{
 \includegraphics[width=0.99\columnwidth]{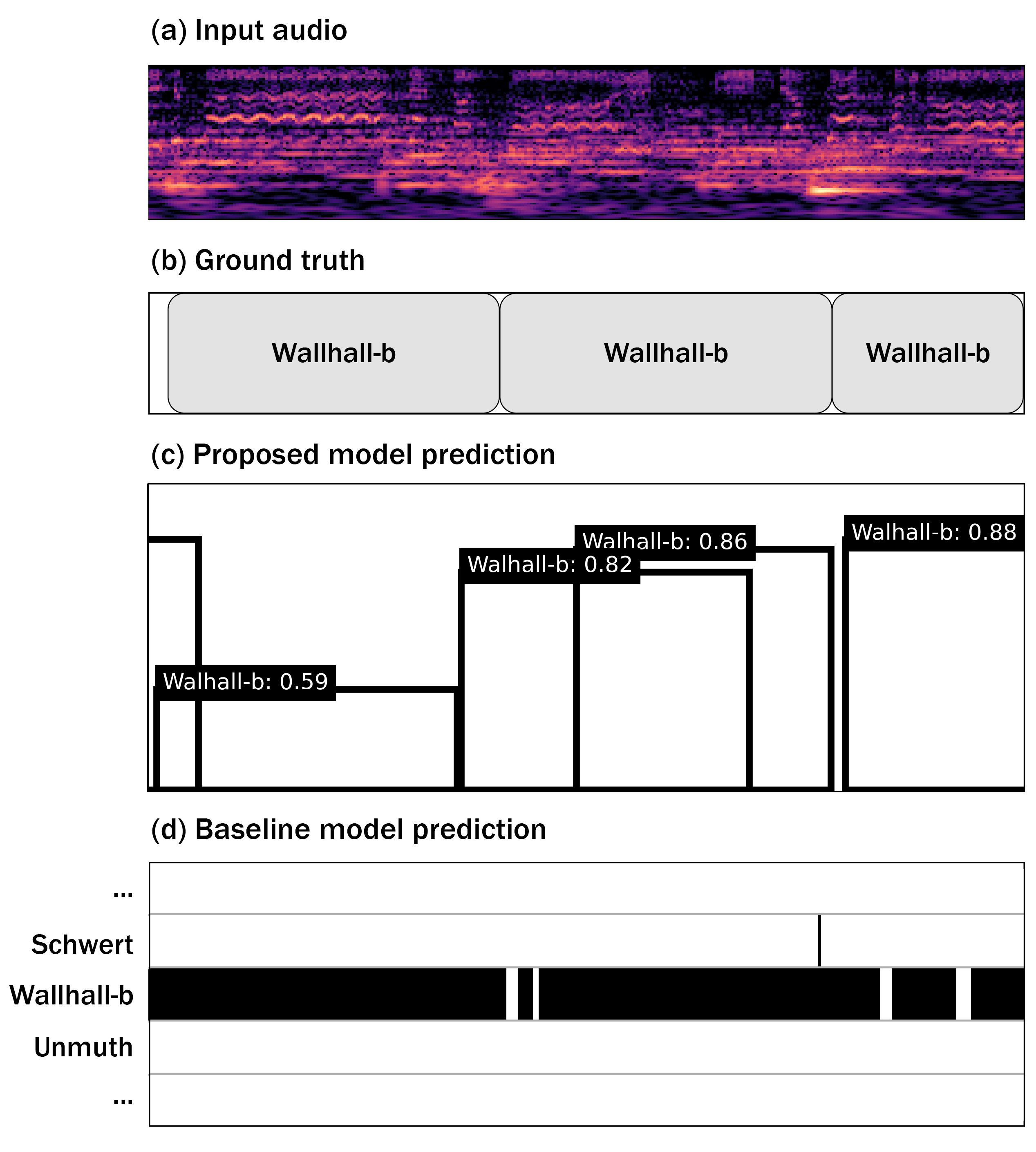}}
 \caption{Predictions from the proposed(c) and baseline(d) models. The heights of the bounding boxes in (c) are relative to confidence scores and do not relate to the frequency dimension.}
 \label{fig:prediction}
\end{figure}

\subsection{Results}
For training, we adapt the multi-part loss function used in YOLOv3\cite{redmon2018yolov3} for single-scale, one-dimensional boundary regression. We use random pitch shift of \(\pm 3\) semitones for augmentation and train for a maximum of 120 epochs with early stopping.

During evaluation, we use non-maximum suppression (NMS) to filter out redundant or irrelevant predictions, and perform a grid search over the confidence and NMS IoU thresholds to maximize F\textsubscript{1} scores for each leitmotif class. We compare the proposed model against a CNN-based baseline model that outputs frame-wise multi-class probabilities, for which we also perform a grid search over the probability threshold.

Table~\ref{tab:results} shows the evaluation results of both models. It should be noted, however, that the metrics from the two models cannot be directly compared, due to the stark difference in how they are calculated.

Figure~\ref{fig:prediction} shows an example from the drama \textit{Rheingold}, where the \textit{Walhall} motif appears repeatedly. While the proposed bounding box regression model and the frame-wise baseline model both show reasonable prediction, the proposed model is able to predict the exact boundaries of each occurrence while the baseline outputs a long consecutive activation of the motif.

We also train and test our model in an "act split" setting, where the training and test set consist of different acts of the same recording version. However, the model fails to make meaningful predictions, similar to the report of the previous study~\cite{krause2020classifying}.

\section{Conclusion}
While our proposed model still has clear limitations, it demonstrates performance comparable to frame-wise prediction methods, while also offering the capability of precisely detecting motif boundaries in repetitive occurrences. This enables accurate quantification of motif appearances, which has potential use in musicological analysis.
We hope that our work draws attention to this intriguing and challenging task of leitmotif detection, stimulating further research and inspiring novel approaches in the field. 

\section{Acknowledgement}
This work was supported by the Ministry of Education of the Republic of Korea and the National Research Foundation of Korea (NRF-2024S1A5C3A03046168).

\bibliography{ISMIR2024_lbd}

%
%
%
%
%

\end{document}